\documentclass[twocolumn,preprintnumbers,aps]{revtex4}

\usepackage{latexsym}
\begin{document}
\bibliographystyle{plain}

\preprint{FERMILAB--CONF--02/049--T}

\title[Next Steps]{Next Steps}



\newcommand{\m}{\hbox{ m}}

\author{Chris Quigg}
\email[]{quigg@fnal.gov}
\thanks{Fermilab is operated by Universities Research Association Inc.
under Contract No.  DE-AC02-76CH03000 with the United States Department
of Energy.}
\affiliation{Fermi National Accelerator Laboratory \\ P.O. Box 500, Batavia IL 60510 USA}


\date{\today}

\begin{abstract}
Closing talk at \textit{Snowmass 2001:} a summer study on the future of particle physics.
\end{abstract}

\maketitle

\section{Thanks!}
We've come to the end of three weeks of intense work, and we have
shattered all precedents for participation in a Snowmass summer study.
We have numbered more than 1200, including more than eighty students,
more than two hundred ``young physicists,'' and more than two hundred
colleagues from outside the United States.  On behalf of the American
Physical Society and its Division of Particles \& Fields and Division of
Physics of Beams, I thank you for coming to Snowmass in such great
numbers, and with such optimism and appetite for hard work.

Throughout the planning and execution of \textit{Snowmass 2001,} we have
benefited from the wise counsel and manifold support of the leaders of the
American Physical Society. I'd particularly like to acknowledge the r\^{o}les 
of Executive Officer Judy Franz, President George Trilling, and Past President
Jim Langer.

It is also my great pleasure to express on your behalf our thanks and
admiration to the people who made our work possible. You may know that
we said we would define 500 participants as a successful summer study,
we planned for about 800, and the people who make things happen have
had to provide for the needs of 1200. I'd like to begin by saluting our
friends in the \textit{Snowmass 2001} conference office: Cindy Arnold, Jody
Federwitz, Ray Fonseca, Michelle Gleason, Carolyn James, Wanda Newby,
Marilyn Paul, Barb Perington, Patti Poole, Cynthia Sazama, Marilyn
Smith, and Suzanne Weber; as well as Shefali Kubavat and Carol Kuc of
Complete Conference Coordinators. Computing support, including the very
helpful wireless network here in the Conference Center, was provided by
Chuck Andrews, John Bellendir, Alden Clifford, Larry English, Steve
Fry, Rick Hill, Kip Kippenham, Andy Rader, Cameron Smith, John Urish,
and Jerry Zimmerman. The streaming-video archive of plenary sessions
and other important events was made possible by Al Johnson, Jim Shultz,
and Fred Ullrich.

All of the work behind the scenes was coordinated by the Local Organizing
Committee led by Jeff Appel. I'm personally grateful to Jeff not only for
making all the pieces fit together, but also for functioning with unfailing grace
and impeccable judgment as a chief-of-staff throughout the workshop. For all of
us, I want to thank Mike Witherell for making available to the summer
study Fermilab's remarkable logistical support. I'm also grateful to Jonathan
Dorfan for making SLAC's resources available to produce the \textit{Proceedings,}
and to Norman Graf for agreeing to serve as Editor. While \textit{Snowmass 2001} is a
creation of the DPF and DPB, it could not have happened without the 
leadership of our great laboratories.

It's a great pleasure to recognize again the marvelous contributions of
my colleagues on the \textit{Snowmass 2001} Organizing Committee, and to thank
them for their wisdom, hard work, and community spirit: Ron Davidson
(PPPL, co-chair), Alex Chao (SLAC), Alex Dragt (Maryland), Gerry Dugan
(Cornell), Norbert Holtkamp (SNS), Chan Joshi (UCLA), Thomas Roser
(BNL), Ron Ruth (SLAC), John Seeman (SLAC), and Jim Strait (Fermilab)
from the DPB side; and Sally Dawson (BNL), Paul Grannis (Stony Brook),
David Gross (ITP/UCSB), Joe Lykken (Fermilab), Hitoshi Murayama
(Berkeley), Ren\'e Ong (UCLA), Natalie Roe (LBNL), Heidi Schellman
(Northwestern), and Maria Spiropulu (Chicago) from the DPF side.  

It's also my happy task to thank the working group convenors for the
extraordinarily rich program we've experienced at \textit{Snowmass 2001}:

\noindent {Accelerator Working Groups.}
\textsf{M1: Muon-Based Systems.} Kirk McDonald (Princeton), Andrew
Sessler (LBNL); \textsf{M2: $e^{+}e^{-}$ Circular Colliders.} K. Oide
(KEK), J. Seeman (SLAC), S. Henderson (Cornell); \textsf{M3: Linear
Colliders.} R. Brinkman (DESY), N. Toge (KEK), T. Raubenheimer (SLAC);
\textsf{M4: Hadron Colliders.} S. Peggs (BNL), M. Syphers (Fermilab);
\textsf{M5: Lepton-Hadron Colliders.} I. Ben-Zvi (BNL), G. Hoffstaetter
(DESY); \textsf{M6: High-Intensity Proton Sources.} W. Chou (Fermilab),
J. Wei (BNL).

\noindent {Technology Working Groups.}
\textsf{T1: Interaction Regions.} T. Markiewicz (SLAC), F. Pilat (BNL);
\textsf{T2: Magnet Technology.} S. Gourlay (LBNL), V. Kashikan
(Fermilab); \textsf{T3: RF Technology.} C. Adolphsen (SLAC), N.
Holtkamp (SNS), H. Padamsee (Cornell); \textsf{T4: Particle Sources.}
J. Sheppard (SLAC), N. Mokhov (Fermilab), S. Werkema (Fermilab);
\textsf{T5: Beam Dynamics.} M. Blaskiewicz (BNL), K.-J. Kim (Argonne),
S. Y. Lee (Indiana); \textsf{T6: Environmental Control.} W. Bialowons
(DESY), C. Laughton (Fermilab), A. Seryi (SLAC); \textsf{T7:
High-Performance Computing.} K. Ko (SLAC), N. Ryne (Los Alamos);
\textsf{T8: Advanced Acceleration Techniques.} C. Joshi (UCLA), P.
Sprangle (NRL); \textsf{T9: Diagnostics.} R. Pasquinelli (Fermilab), M.
Ross (SLAC).

\noindent {Physics Working Groups.}
\textsf{P1: Electroweak Symmetry Breaking.} Marcela Carena (Fermilab),
David Gerdes (Michigan), Howard Haber (Santa Cruz), Andr\'{e} Turcot
(Brookhaven),  Peter Zerwas (DESY); \textsf{P2: Flavor Physics.}
Bel\'{e}n Gavela (Madrid), Boris Kayser (NSF), Clark McGrew (Stony
Brook), Patricia Rankin (Colorado); \textsf{P3: Scales beyond 1 TeV.}
Michael Dine (Santa Cruz), JoAnne Hewett (SLAC), Greg Landsberg
(Brown), David Miller (UCL); \textsf{P4: Astro/Cosmo/Particle Physics.}
Dan Akerib (Case-Western Reserve), Sean Carroll (Chicago), Mark
Kamionkowski (Caltech), Steve Ritz (NASA/Goddard); \textsf{P5: QCD \&
Strong Interactions.} Brenna Flaugher (Fermilab), Ed Kinney (Colorado),
Paul Mackenzie (Fermilab), George Sterman (Stony Brook).

\noindent {Experimental Approaches Working Groups.}
\textsf{E1: Neutrino Factories \& Muon Colliders.} Vernon Barger
(Wisconsin), Debbie Harris (Fermilab), Yoshi Kuno (Osaka), Mike Zeller
(Yale, HMO); \textsf{E2: $e^{+}e^{-}$ Colliders below the $Z$.} Gustavo
Burdman (BU), Ian Shipsey (Purdue), Hitoshi Yamamoto (Hawaii), Joel
Butler (Fermilab, HMO); \textsf{E3: Linear Colliders.} Marco Battaglia
(CERN), John Jaros (SLAC), James Wells (Davis),Ian Hinchliffe (LBNL,
HMO); \textsf{E4: Hadron \& Lepton-Hadron Colliders.} Uli Baur
(Buffalo), Raymond Brock (Michigan State), John Parsons (Columbia),
Bill Marciano (Brookhaven, HMO); \textsf{E5: Fixed-Target Experiments.}
Krishna Kumar (UMass), Ron Ray (Fermilab), Paul Reimer (Argonne), Mark
Strovink (Berkeley, HMO); \textsf{E6: Astro/Cosmo/Particle
Experiments.} Kevin Lesko (LBNL), Suzanne Staggs (Princeton), Tim McKay
(Michigan), Harry Nelson (UCSB, HMO); \textsf{E7: Particle Physics \&
Technology.} Stephan Lammel (Fermilab), Wesley Smith (Wisconsin).
	
Thanks as well to the IEEE/NPSS Committee for Technology Emphasis:
Bruce C. Brown, Matthew A. Allen, William M. Bugg, Peter Clout, John E.
Elias, Erik Heijne, Thomas Katsouleas, Ray S. Larsen, Patrick Le Du,
Alan Todd, Craig L. Woody, and to the thirty-eight presenters of
lectures and courses.

The exceptional program of engagement with the public here at Snowmass
2001 was constructed by the Outreach Coordinating Committee made up of
Elizabeth Simmons (chair), Marge Bardeen, Martin Berz, Bill Frazer,
Evalyn Gates, Joey Huston, Ronen Mir, Mel Month, Helen Quinn, Deborah
Roudebush, Greg Snow, Ken Taylor, and Jeff Wilkes, with staff support
from Melissa Clayton. More than a hundred of you took part in Science
Weekend and the other outreach activities, which were notable for the
great joy they brought both to the public and to the physicist
participants. I'd especially like to recognize the contributions of
Mayda Velasco (who animated \textit{La Noche de la Ciencia}), Marko
Popovic, Kevin Lynch, Shreyas Bhat, Lawrence Krauss, Leon Lederman, and Liz Quigg.
Support for the \textit{Snowmass 2001} outreach and education program was
provided by the Division of Particles and FieldsÊ and the Division of
Physics of Beams of the American Physical Society, the National Science
Foundation, the U.S. Department of Energy, the ATLAS and CMS
Collaborations, Universities Research Association, the IEEE Council on
SuperConductivity, Maggie and Nick DeWolf, and by other public-spirited
individuals and organizations. We appreciate very much the cooperation
of Snowmass Village for giving us the run of the mall for Science
Weekend and for other courtesies, and the collaboration of Explore
Booksellers of Aspen, the Aspen Public Schools, Camp Snowmass (Sue
Way), the Science Outreach Center of Carbondale (Linda Froning), and
the Carbondale Community School.

Also making \textit{Snowmass 2001} a public event---in the Roaring Fork Valley
and around the world---was our full-time (which came close to 24/7) Press Room.
I thank Judy Jackson, Mike Perricone, Kurt Riesselman, and Mieke van
den Bergen (all of Fermilab) for all their contributions---visible and 
behind the scenes---to the success of our summer study.

Thanks again to our sponsors, the United States Department of Energy,
the National Science Foundation, and NASA, and to the ten laboratories
engaged in particle physics research in the United States who have
supported our cause: Argonne National Lab, Berkeley Lab, Brookhaven
National Lab / Brookhaven Science Associates, Cornell University / LNS
/ Wilson Synchrotron Lab, Fermilab / Universities Research Association,
Jefferson Laboratory / SURA, Lawrence Livermore National Laboratory,
Los Alamos National Laboratory, Oak Ridge National Lab / Spallation
Neutron Source, Stanford Linear Accelerator Center / Stanford
University. 

\section{Wonderful Things \protect\\ Have Been Happening Here}
We have rediscovered our community and our sense of common destiny.
We have celebrated the astonishing progress and remarkable promise of
particle physics, broadly understood. 

\textit{No one should miss the
conclusion that ours is a community on the move, worldwide.}

We have taken pleasure in the inventiveness and careful thought of our
colleagues who dream, design, and build accelerators and the components
that make them possible. We have mixed; we have engaged with each
other's aspirations and significantly advanced a number of ideas. Every
large future accelerator project is taken more seriously, and valued
more highly, than it was when we assembled three weeks ago. That
outcome is a testimony both to the superb work being done by our
colleagues who create accelerator possibilities and to the generous and
open spirit all of you brought to Snowmass.

Our conception of the scientific landscape has been enriched by the
high level of participation by our astro/cosmo/particle colleagues,
including those from communities that have not traditionally 
identified with  particle physics. 

We have relished the global reach of our science. 
We profited from the contributions---formal and informal---of international
laboratory directors Alessandro Bettini (Gran Sasso), Luciano Maiani
(CERN), Alexander Skrinsky (BINP), Hirotaka Sugawara (KEK), and Albrecht
Wagner (DESY), as well as the spirited participation of the leaders of
U.S. laboratories, Jonathan Dorfan (SLAC), Tom Kirk (BNL), Maury Tigner
(Cornell), and Mike Witherell (Fermilab). Ian Corbett brought important
insights into the work of the Global Science Forum, while Ferdi Willeke
reported on the work of the ICFA Committees exploring the idea of a
Global Accelerator Network. Lorenzo Fo\`{a} and Satoru Yamashita presented
conclusions of the European and Japanese planning exercises.

We have enjoyed an atmosphere of excellence and optimism and savored the
inspiring promise of youth. 
Young physicists plunged into the activities of the working groups and
also created a vigorous program of their own, which culminated in the
Young Physicists Forum attended by many Snowmass participants of all
ages. A Young Particle Physicists organization is taking shape. For the
latest word, see the web site at \url{http://ypp.hep.net}; results of the
survey carried out by the young physicists can be found at
\url{http://ypp.hep.net/ypp_survey.html}. 

Other special events included half-day ``teach-ins'' on opportunities in
accelerator research and development and on the new world of
astro/cosmo/particle physics, and communications workshops that brought
together science writers, Washington communicators, public affairs professional
from many laboratories in the US and abroad, and physicists. The
HEPAP subpanel on long-range planning, led by Jonathan Bagger and Barry
Barish, participated actively in the work of \textit{Snowmass 2001}, and used
the workshop to solicit and receive input of many
kinds from the community. The NAS/NRC Committee on Physics of the
Universe, led by Mike Turner, also met at Snowmass, as did the High
Energy Physics Advisory Panel.

The main products of \textit{Snowmass 2001} will, as always, be the individual
contributions and the working group summaries to appear in the
Proceedings. This year, we undertook two special projects as well. The
Division of Particles and Fields has produced a twenty-page illustrated
survey of the grand themes and aspirations of particle physics, called
\textit{Quarks Unbound,} which will soon appear in the mailboxes of all
DPF members. \textit{Quarks Unbound} was written by science writer
Sharon Butler, based on her interviews and discussions with many
members of the community. We owe a special debt of gratitude to two
members of the Snowmass Organizing Committee, Joe Lykken and Maria
Spiropulu, for their dedication to the project. We will be using
\textit{Quarks Unbound} widely to communicate the excitement and
promise of particle physics. The Division of Physics of Beams has used
the work at Snowmass to create a cogent blueprint for future
accelerator research and development (see
\url{http://www.hep.anl.gov/pvs/dpb/Snowmass.pdf}). All this is only
the beginning; we planted many seeds at \textit{Snowmass 2001}, and we will be
harvesting their fruits for many years.

\section{Reviewing the Goals \protect\\ for Snowmass 2001}
We have made decisive progress on all the goals I set out three weeks
ago, and I trust that you have discovered many new things to think
about and to work on in the months and years to come.

We have made an excellent start toward surveying our aspirations for
particle physics over 30 years by educating ourselves about the full
range of possibilities before us.  We were aided immeasurably in the
Experimental Approaches working groups by the ``High-Minded
Outsiders,'' who served as friendly skeptics, probing and strengthening
arguments. Our explorations were aided as well by the presence of
string theorists who helped us look beyond our immediate goals to
contemplate the shape of a more complete, more ambitious theoretical
framework.  On the machine front, the work carried out at Snowmass has
help us to  understand the investment we must make (financial and human
capital) to bring the most promising lines to maturity.  

We need to continue the conversation about the degree of scientific
diversity, including scale diversity, that we need to build a healthy
and productive future for particle physics. 

We must continue to work toward articulating a comprehensive vision of
particle physics (and the sciences it touches) to make our case
effectively to ourselves, to other scientists, and to society at large.

Three weeks ago, I observed that ``the moment is upon us to probe,
shape, and judge the idea of a linear collider as a possible next big
step for particle physics. \textit{Evaluating a linear collider and
working to define a scientifically rich, technically sound, fiscally
responsible plan is a homework problem for the entire community.} 
Everyone must come to an informed judgment.'' 

At \textit{Snowmass 2001}, a widespread feeling has emerged that the world
community should move urgently to construct a TeV-scale linear collider
as an international project. \emph{These are ambitious machines and
significant challenges remain: we  must be certain of the costs and we
must take the measure of technical risks. A phase change is needed to
complete the design and development promptly.}  

In the United States, another phase change is needed \textit{soon} in
the commitment of experimental physicists to the linear collider
program. A few people have done valuable work, but outside the US,
\textit{many more people} have done much more comprehensive work. US
participation in a linear collider will not be decisive without the
engagement of a large and energetic cadre of superb experimenters to
hone the physics case, participate in parameter choices, and work
side-by-side with the machine builders. \textit{If you wait, it will
not happen!} It is also time for closer cooperation among physicists in
different regions on linear collider issues: to coordinate R\&D, to
develop a unified physics document, and to make the scientific case to
the governments of the world---perhaps it is the moment to form an
International Linear Collider Users Group?

\section{When you go home \ldots}

Continue to think about what you have heard and done at Snowmass.
	
Talk with your particle physics colleagues about what you have seen and
heard and done here.  Arrange seminars to share the \textit{Snowmass 2001}
experience with all your students and colleagues.
	
Write your advice to the HEPAP subpanel; If you wrote long ago, reread
your letter to see how your thinking has evolved.
	
Talk with your colleagues in other fields of physics and astronomy
about \textit{Snowmass 2001.} Share your enthusiasm! Give a colloquium early in
the school year about the future of particle physics.
	
Talk with your colleagues in other fields about their excitement and
aspirations.  Help your students appreciate the exciting futures all
across physics and astronomy.
	
We've heard at \textit{Snowmass 2001} from many gifted, articulate, and
inspiring colleagues:  Invite them to visit your department.  Hire them!
	
No department is whole without some presence in experimental particle
physics, particle theory that engages with experiment, accelerator
physics, astro/cosmo/particle physics, string theory (to speak only of
our immediate neighborhood). Make your institution stronger, and you strengthen
our science.

\section{Mike Holland's \protect\\ ``Insulting'' Questions}
Members of Congress, Congressional Staff, and White House Staff are
busy people.  Their first response to any request is, ``No.''
\textit{If you go away and never come back, they gave the right
answer.} 

Their second response to any request is, ``You don't have a plan.''
\textit{If you go away and never come back, they gave the right answer.}
	
It is not unknown for our friends to ask hostile questions in hopes of
learning whether we have answers they can use when they are asked
hostile questions.
	
Some people in Washington and some scientists do not have our interests
at heart.  We must not let them seize the agenda and frame the debate,
but when they ask easy questions for nefarious purposes, we should leap
to give compelling answers. So when someone throws you a fat pitch,
asking ``Does particle physics require accelerators?'' or ``What is the
value of particle physics to other sciences and to society?'' do not
squander time and adrenaline fulminating about the injustice, just
smack the ball over the fence.

\section{Ask More of the \protect\\ United States Government}
The difficult environment for funding the physical sciences reflects, I
believe, a deeper problem in our society. For more than a decade, the
will to join together and undertake challenging and important causes
for the general good has been too little in evidence. \textit{This is
an aberration in American history, and we must change it.}
	
In a time of unparalleled prosperity, every section of every
appropriations bill seems to begin, ``Because of severe budgetary
constraints \ldots'' We are still waiting for the peace dividend.

What can we do? We must demand better!  The public believes in science
and exploration, and we are asking questions that engage the public's 
imagination. We must help people in government and the media to
understand this.

Basic research (and not only particle physics) is a superb investment
on many levels. Don't be timid (but be sensible). Many people are
dining out on the World Wide Web, an unprogrammed dividend of a tiny
fraction of the world's investment in particle physics. If people want
to count beans, we must insist that they count \emph{all} the beans.

Like every individual, every nation must decide what constitutes a
meaningful life.  Share your passions and your dreams, and lift the
eyes of those who govern!

\section{Onward and Upward!}
Making a new world is not accomplished in three weeks or three months
or three years, but we have made a wonderful start here at
\textit{Snowmass 2001.} Your passion, energy, creativity, and
commitment truly did begin to change the world. Thank you!

\end{document}